# Role of Interlayer Coupling in the Second Harmonic Generation of Bilayer Transition-metal Dichalcogenides

Xiao Jiang[1], Lei Kang[2], and Bing Huang*[1,3]
1. Beijing Computational Science Research Center, Beijing 100193, China
2. Technical Institute of Physics and Chemistry, Chinese Academy of Sciences, Beijing 100190, China
3. Department of Physics, Beijing Normal University, Beijing 100875, China

E-mail: Bing.Huang@csrc.ac.cn

Little is known about the role of weak interlayer coupling in the second harmonic generation (SHG) effects of two-dimensional van der Waals (vdW) systems. In this article, taking homo-bilayer $MoS_2/MoS_2$ and hetero-bilayer $MoS_2/MoSe_2$ as typical examples, we have systemically investigated their SHG susceptibilities $\chi^{(2)}$ as a function of interlayer hopping strength ($t_{int}$) using first-principles calculations. For the $\chi^{(2)}_{yyy}(0;0,0)$ of both $MoS_2/MoS_2$ and $MoS_2/MoSe_2$, although the increase of $t_{int}$ can increase the intensities of interlayer optical transitions (IOT), the increased band repulsion around $\Gamma$ point can eventually decrease their $\chi^{(2)}_{yyy}(0;0,0)$ values; the larger the $t_{int}$, the smaller the $\chi^{(2)}_{yyy}(0;0,0)$. For the $|\chi^{(2)}_{yyy}(-2\omega;\omega,\omega)|$ spectra of $MoS_2/MoSe_2$ in the low photon-energy region, opposite to the $MoS_2/MoS_2$, their peak values are very sensitive to the variable $t_{int}$, due to the strong $t_{int}$-dependent IOT dominating in the band edge; the larger the $t_{int}$, the larger the $|\chi^{(2)}_{yyy}(-2\omega;\omega,\omega)|$. For the $|\chi^{(2)}_{yyy}(-2\omega;\omega,\omega)|$ of $MoS_2/MoS_2$ in the high photon-energy region, comparing to the $MoS_2/MoSe_2$, their peak values will decrease in a much more noticeable way as the $t_{int}$ increases, due to the larger reduction of band nesting effect. Our study not only can successfully explain the puzzling experimental observations for the different SHG responses in different bilayer transition-metal dichalcogenides under variable $t_{int}$, but also may provide a general understanding for designing controllable the SHG effects in the vdW systems.

# I. INTRODUCTION

Two-dimensional (2D) layered materials are promising to extend the functionalities from bulk to nanoscale thickness [1-3]. Among them, the noncentrosymmetric materials with large second-order optical susceptibilities, *e.g.*, optical second harmonic generation (SHG) [4-6], linear electro-optic or Pockels effects, sum frequency generation, and optical rectification [7], are highly valuable for various practical applications, *e.g.*, electro-optic modulators, frequency conversion, structure and magnetic orders probing [8-10]. Recently, many monolayer 2D materials are discovered to exhibit large second or third harmonic generations, including several transition-metal dichalcogenides (TMDs) (e.g., $MoS_2$ [11-15], $MoSe_2$ [16-18], $WS_2$ [19,20], and $WSe_2$ [21]), hexagonal BN [22], InSe [23] and GaSe [24]. Interestingly, despite their monolayer thickness, their effective SHG responses are comparable to many widely used bulk NLO materials [13]. Benefiting from their nanoscale thickness, these 2D systems are promising for on-chip applications without the requirement of phase-matching conditions [8,25].

Van de Waals (vdW) engineering has been widely applied to manipulate the electronic and optoelectronic properties of 2D materials. For example, the vdW homo-bilayer (e.g., $MoS_2/MoS_2$ [26,27] and bilayer graphene [28-30]) and hetero-bilayer (e.g., $MoS_2/MoSe_2$ [26,31-33] and graphene/*h*-BN [34]) structures can exhibit fundamentally different electronic and optical properties compared to their individual components [27], due to the existence of weak interlayer coupling effects. In practice, the interlayer coupling strength of these vdW bilayers can be either enhanced by the external pressure/post-annealing [26,35-37] or decreased by the wide bandgap thin-films inserted/dry transferred [26]. Surprisingly, it is found that depending on the different types of TMDs, the SHG responses in these systems can be either sensitive (e.g., in hetero-bilayers) or insensitive (e.g., in homo-bilayers) to the variable interlayer coupling strengths [26]. However, until now, it is still unclear about the physical origin behind these puzzling experimental observations, preventing the design of controllable SHG effects in the 2D systems.

In this article, taking homo-bilayer $MoS_2/MoS_2$ and hetero-bilayer $MoS_2/MoSe_2$ as the typical examples, we have systemically investigated the SHG susceptibilities $\chi^{(2)}$ of these bilayer TMDs as a function of interlayer hopping strength ($t_{int}$) using first-principles Wannier function calculations. Interestingly, for the $\chi^{(2)}_{yyy}(0;0,0)$, it is found that although the increase of $t_{int}$ can increase the interlayer optical transitions, the increased band repulsion around the $\Gamma$ point in the band structures of both $MoS_2/MoS_2$ and $MoS_2/MoSe_2$ can largely decrease their $\chi^{(2)}_{yyy}(0;0,0)$ values. The larger the $t_{int}$, the smaller the $\chi^{(2)}_{yyy}(0;0,0)$. For the $|\chi^{(2)}_{yyy}(-2\omega;\omega,\omega)|$ spectra of bilayer $MoS_2/MoS_2$, it is found that the intensities of low photon-energy peaks are insensitive to the variable $t_{int}$, but the intensities of high photon-energy peaks will gradually decrease as the $t_{int}$ increases, due to the reduced band nesting effect. Differing from the bilayer $MoS_2/MoS_2$, it is found that the intensities of low photon-energy peaks in the $|\chi^{(2)}_{yyy}(-2\omega;\omega,\omega)|$ spectra of $MoS_2/MoSe_2$ can significantly increase as the $t_{int}$ increases, due to the strongly enhanced band-edge optical transitions; but the intensities of high photon-energy peaks will decrease as the $t_{int}$ increases, which is less noticeable than that in the $MoS_2/MoS_2$. Our findings not only can well explain the puzzling experimental observations for the different SHG responses in the homo- and hetero-bilayer TMDs under variable interlayer coupling strengths, but also could be important for designing tunable SHG effects in other 2D vdW systems.

## II. METHODS

### 2.1 SHG calculations

The SHG susceptibility tensor $\chi^{abc}_{total}(-2\omega;\omega,\omega)$ can be calculated as [38,39]:

$$\chi^{abc}_{total}(-2\omega;\omega,\omega) = \chi^{abc}_e + \chi^{abc}_i \quad (1).$$

In Eq. (1), the $\chi^{abc}_e$ and $\chi^{abc}_i$ terms originate from the interband contribution and the mixed interband and intraband contributions, respectively, which can be written as:

$$\chi^{abc}_e(-2\omega;\omega,\omega) = \frac{e^3}{\hbar^2\Omega}\sum_{nml,k}\frac{r^a_{nm}\{r^b_{ml}r^c_{ln}\}}{(\omega_{ln}-\omega_{ml})}\left[\frac{2f_{nm}}{\omega_{mn}-2\omega} + \frac{f_{ln}}{\omega_{ln}-\omega} + \frac{f_{ml}}{\omega_{ml}-\omega}\right] \quad (2)$$

and

$$\chi_i^{abc}(-2\omega;\omega,\omega) = \frac{i}{2}\frac{e^3}{\hbar^2\Omega}\sum_{nml,k}\frac{2}{\omega_{mn}(\omega_{mn}-2\omega)}r_{nm}^a(r_{nm;c}^b + r_{mn;b}^c) +$$

$$\frac{1}{\omega_{mn}(\omega_{mn}-\omega)}(r_{nm;c}^a r_{mn}^b + r_{nm;b}^a r_{mn}^c) + \frac{1}{\omega_{mn}^2}\left(\frac{1}{\omega_{mn}-\omega} - \frac{4}{\omega_{mn}-2\omega}\right)r_{nm}^a(r_{nm}^b\Delta_{mn}^c + r_{mn}^c\Delta_{mn}^b) -$$

$$\frac{1}{2\omega_{mn}(\omega_{mn}-\omega)}(r_{nm;a}^b r_{mn}^c + r_{nm;a}^c r_{mn}^b) \quad (3).$$

In Eqs. (2) and (3), the $f_{nm} = f_n - f_m$, $\omega_{nm} = \omega_n - \omega_m$, and $\Delta_{nm} = v_{nn} - v_{mm}$ are the Fermi distribution function difference, energy difference, and electron group velocity difference between the $n$th and $m$th bands, respectively. The $r_{nm}$ is the position operator $r_{nm} = \frac{v_{nm}}{i\omega_{nm}}$, representing the optical transition dipole moments (TDMs) between $n$th and $m$th bands. The $\Omega$ is the volume of unit cell, and $\{r_{ml}^b r_{ln}^c\}$ is defined as $1/2(r_{nl}^b r_{lm}^a + r_{nl}^a r_{lm}^b)$. The $r_{nm;a}^b$ is the generalized derivative of the coordinate operator in momentum space with the form of

$$r_{nm;b}^a = \frac{i}{\omega_{nm}}\left[\frac{v_{nm}^a\Delta_{nm}^b + v_{nm}^b\Delta_{nm}^a}{\omega_{nm}} - W_{nm}^{ab} + \sum_{p\neq n,m}\left(\frac{v_{np}^a v_{pm}^b}{\omega_{pm}} - \frac{v_{np}^b v_{pm}^a}{\omega_{np}}\right)\right] \quad (4),$$

where $W_{nm}^{ab} = \langle n|\partial_{k_a}\partial_{k_b}H|m\rangle$ [40,41]. In practical calculations, the frequency ω in the denominator of SHG susceptibility tensor has a small imaginary smearing factor δ (ω → ω + iδ), and δ = 0.05 eV is used in this study (see **Fig. S1a** [42] for the test about δ).

At zero frequency limit (*ω*=0), the value of $\chi_{total}^{abc}(-2\omega;\omega,\omega)$, *i.e.*, $\chi_{total}^{abc}(0;0,0)$, can be reduced to:

$$\chi_e^{abc} = \frac{e^3}{\hbar^2\Omega}\sum_{nml,k}\frac{r_{nm}^a\{r_{ml}^b r_{ln}^c\}}{\omega_{nm}\omega_{ml}\omega_{ln}}[\omega_n f_{ml} + \omega_m f_{ln} + \omega_l f_{nm}] \quad (5)$$

and

$$\chi_i^{abc} = \frac{i}{4}\frac{e^3}{\hbar^2\Omega}\sum_{nml,k}\frac{f_{nm}}{\omega_{mn}^2}[r_{nm}^a(r_{nm;c}^b + r_{mn;b}^c) + r_{nm}^b(r_{mn;c}^a + r_{mn;a}^c) + r_{nm}^c(r_{mn;b}^a + r_{mn;a}^b)] \quad (6).$$

The $\chi_{total}^{abc}(0;0,0)$ is an important physical quantity for evaluating the intrinsic SHG properties of a system.

The general expression of SHG susceptibility $\chi_{abc}^{(2)}(-2\omega;\omega,\omega)$ can be simplified as:

$$\chi^{(2)}_{abc}(-2\omega;\omega,\omega) = \sum_{nml,\mathbf{k}} A_{nml}(\mathbf{k},\omega) \quad (7).$$

To obtain the band-resolved SHG susceptibility tensor, it can be decomposed into the individual bands by partially summing only two out of all the three bands indices [43]. For example, the resolved SHG susceptibility for the *nth* band can be written as:

$$\chi^{(2)}_{abc,n\mathbf{k}}(-2\omega;\omega,\omega) = \sum_{ml} A_{nml}(\mathbf{k},\omega) \quad (8).$$

Moreover, the SHG susceptibility tensor can be projected to the full Brillouin zone by summing over all the three bands indices [44]:

$$\chi^{(2)}_{abc,\mathbf{k}}(-2\omega;\omega,\omega) = \sum_{nml} A_{nml}(\mathbf{k},\omega) \quad (9).$$

**2.2 First-principles calculations**

The first-principles based density functional theory (DFT) calculations are performed using Vienna Ab initio Simulation Package (VASP) [45]. The pseudopotentials of the projector augmented wave (PAW) type are adopted to describe the interaction between valence electrons and ionic cores [46]. The exchange-correlation energy is treated within the generalized gradient approximation (GGA) as parametrized by Perdew, Burke, and Ernzerhof (PBE) [47]. The energy cutoff of 500 eV and Monkhorst–Pack *k*-point meshes of $15 \times 15 \times 1$ in the 2D Brillouin zone are adopted in our study, which can be used to obtain the converged results. The convergence criteria for force and energy are set to be $1\times10^{-3}$ eV and $1\times10^{-5}$ eV/Å, respectively. The vacuum layer is set to 30 Å, which is sufficiently large to avoid the artificial interlayer coupling between the neighboring supercells. The vdW interactions of the 2D systems are treated by the DFT-D3 method [48].

Because of the ill-defined thickness of 2D systems, the calculated SHG susceptibility should be renormalized [49,50]. Similar to the sheet conductance of 2D materials, we define the 2D SHG susceptibility as $\chi^{SHG}_{sheet}(-2\omega;\omega,\omega) = \chi^{SHG}_{bulk}(-2\omega;\omega,\omega) \times L_z$ [48]. Here, $L_z$ the length of the entire unit cell along the vacuum direction [5]. Therefore, based on this definition, the unit of sheet SHG susceptibility is pm$^2$/V.

The Wannier fitting of DFT band structure is implemented within the Wannier90 code [51], in which the Hamiltonian in Wannier basis can be constructed to obtain the $r_{nm}$ for the SHG calculations. The Hamiltonian for bilayer TMDs can be written as

$$\hat{H}^{(2L)} = \sum_k [\hat{\varphi}_1^\dagger(k) H_1^{(1L)} \hat{\varphi}_1(k) + \hat{\varphi}_2^\dagger(k) H_2'^{(1L)} \hat{\varphi}_2(k) + \hat{\varphi}_2^\dagger(k) V_{int}^{(LL)} \hat{\varphi}_1(k) + H.C.] \quad (10)$$

where $H_1^{(1L)}$ and $H_2'^{(1L)}$ are the Hamiltonian of the individual monolayer TMDs, and $\hat{\varphi}_1(k)$ and $\hat{\varphi}_2(k)$ are their corresponding wavefunctions. As shown in Eq. (10), the interlayer interactions in bilayer TMDs can be introduced by $V_{int}^{(LL)}$. In our study, the interlayer hopping strengths $t_{int}$ in $V_{int}^{(LL)}$ can be artificially tuned to different values to simulate the different interlayer coupling strengths. Meanwhile, the different interlayer distances ($d_{int}$) corresponding to the different $t_{int}$ can be calculated via comparing their band structures. In particular, $t_{int}=t_0$ is defined as the interlayer coupling strength of bilayer TMDs at the equilibrium interlayer distance.

## III. Result and Discussion

In Sections 3.1 and 3.2, we will discuss the role of $t_{int}$ in the SHG responses in the homo-bilayer $MoS_2/MoS_2$ systems, respectively. In Sections 3.3 and 3.4, we will discuss the role of $t_{int}$ in the SHG responses in the hetero-bilayer $MoS_2/MoS_2$ systems, respectively. Since the SHG effects can only exist in the systems without inversion symmetry, the bilayer TMD systems with 3R-type are considered, which have been synthesized in the prior experiments [12,26].

### 3.1 $\chi_{yyy}^{(2)}(0;0,0)$ of the homo-bilayer $MoS_2/MoS_2$ as a function of $t_{int}$

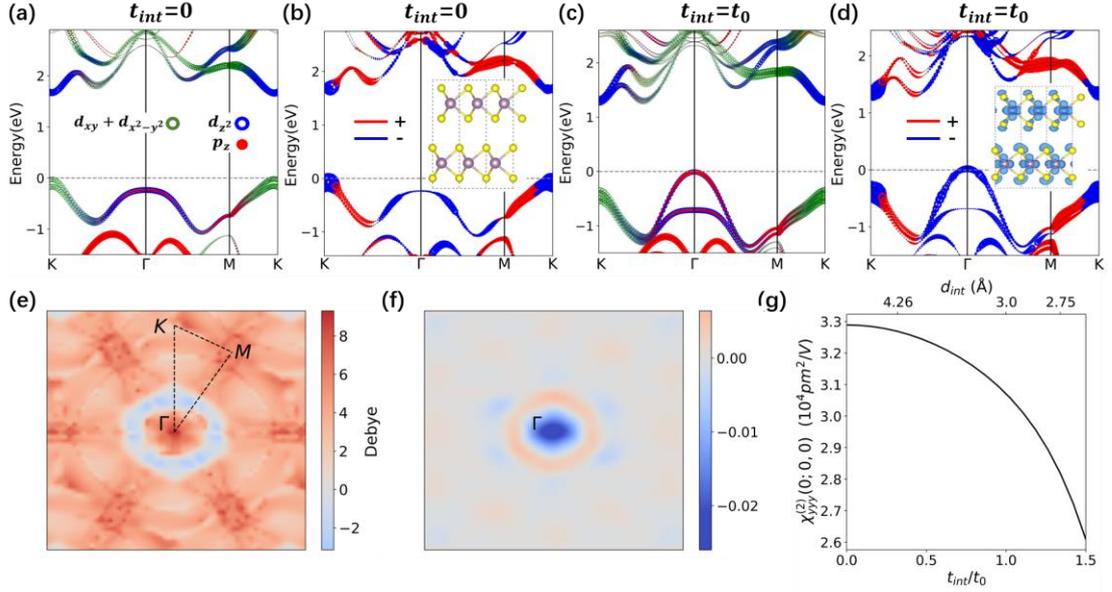

**Fig. 1.** (a) and (b) are orbital-resolved and SHG-weighted band structures of the homo-bilayer MoS$_2$/MoS$_2$ ($t_{int}$=0), respectively. Spin-orbital coupling (SOC) effects are considered in the band structure calculations. Fermi level is set to zero. Inset of (b): Side view of the MoS$_2$/MoS$_2$ in *3R* phase. (c)-(d) are same as (a)-(b) but for the case of MoS$_2$/MoS$_2$ ($t_{int}$=$t_0$). Inset of (d): SHG-weighted charge density of MoS$_2$/MoS$_2$ in the real space. See the enlarged versions of the insets in (b) and (d) in Fig. S2 [42]. (e): Difference of TDMs between the cases of $t_{int}$=0 and $t_{int}$=$t_0$ in the full Brillouin zone (BZ) for the MoS$_2$/MoS$_2$. Here, the plotted TDMs are the sum of all the four optical transition processes between the top two valence bands and the bottom two conduction bands. (f): Difference of SHG projection between the cases of $t_{int}$=0 and $t_{int}$=$t_0$ for the MoS$_2$/MoS$_2$ in the full BZ. (g): Values of SHG susceptibility $\chi^{(2)}_{yyy}(0;0,0)$ for the MoS$_2$/MoS$_2$ as a function of $t_{int}$.

The structure of bilayer MoS$_2$/MoS$_2$ in *3R* phase (*R3m*) is shown as the inset of **Fig. 1b**. For the bilayer MoS$_2$/MoS$_2$ with decoupled interlayer coupling effect, *i.e.*, $t_{int}$=0, it is a direct bandgap ($E_g$) semiconductor (DFT-calculated $E_g$=1.7 eV) with both conduction band minimum (CBM) and valence band maximum (VBM) locating at the *K* point, same as the monolayer MoS$_2$ [15]. Due to the strong spin-orbital coupling (SOC) effect, the double-degenerated CBM or VBM in each MoS$_2$ layer will split into two single ones with opposite spin directions at the *K* point [15]. As shown in **Fig. 1a**, the orbital-resolved band structure shows that the CBM and VBM are mostly contributed by the Mo $d_{z^2}$ and Mo $d_{x^2-y^2} + d_{xy}$ orbitals, respectively.

As shown in **Fig. 1c**, when $t_{int}$ increases from 0 to $t_0$ (the hopping strength at the equilibrium interlayer distance $d_{int}$=3.0 Å), the interlayer coupling effect can significantly modulate the band structure of MoS$_2$/MoS$_2$, especially around the highest-symmetry $\Gamma$ point, inducing a direct-to-indirect bandgap transition (DFT-calculated $E_g$=1.3 eV) [27]. Comparing the cases between $t_{int}$=0 (**Fig. 1a**) and $t_{int}$=$t_0$ (**Fig. 1c**), it is observed that there is a large band repulsion around the $\Gamma$ point at the top of valence band, shifting the VBM to the $\Gamma$ point. Now, the VBM state is contributed by the Mo-$d_{z^2}$ and S-$p_z$ orbitals among the two different layers. Therefore, differing from the simple three-orbitals ($d_{z^2}$, $d_{xy}$, $d_{x^2-y^2}$) tight-binding model for monolayer MoS$_2$ [52], the S $p$ orbitals must be included to capture the low-energy band structure of bilayer MoS$_2$/MoS$_2$ [53]. Accordingly, to accurately describe the interlayer coupling effect and its role in the SHG response, we have constructed the Hamiltonian in Wannier basis using 20 Mo $d$ orbitals and 24 S $p$ orbitals (including the SOC effects) to generate the localized Wannier functions (see **Fig. S1b** [42] for the convergence test of the SHG calculations as a function of energy bands). Overall, the Wannier fitted band structure agrees well with the DFT one (see **Fig. S3** [42]).

When the $t_{int}$ increases, the interlayer wavefunction overlap can increase, which may give rise to the increased optical transition intensities in the system. By calculating the TDMs between top two valence bands and bottom two conduction bands in the full Brillouin zone (BZ) of MoS$_2$/MoS$_2$ under $t_{int}$=0 (**Fig. S4a** [42]) and $t_{int}$=$t_0$ (**Fig. S4b** [42]), we can obtain the difference between them. As shown in **Fig. 1e**, indeed, it is interesting to observe that the optical transition intensities are increasing in the entire BZ except for a small area around the $\Gamma$ point.

Generally, it is expected that increased TDMs may increase the $\chi^{(2)}$. To test this intuition, we have calculated the $\chi^{(2)}$ of MoS$_2$/MoS$_2$. Because of the $D_{3h}$ point group symmetry, there is only one independent SHG response susceptibility tensor elements $\chi^{(2)}_{yyy} = -\chi^{(2)}_{yxx} = -\chi^{(2)}_{xyx} = -\chi^{(2)}_{xxy}$. Surprisingly, it is found that the calculated value of SHG susceptibility $\chi^{(2)}_{yyy}(0;0,0)$ decreases (not increases) from 3.3×10$^4$ to 3.0×10$^4$ pm$^2$/V when $t_{int}$ increases from 0 to $t_0$. The SHG-weighted charge density (inset of **Fig. 1d**) shows that the $\chi_{yyy}^{(2)}(0;0,0)$ is mainly contributed by Mo $d_{x^2-y^2} + d_{xy}$, Mo $d_{z^2}$, and S $p_z$ orbitals, regardless of the specific value of $t_{int}$.

It is interesting to further understand this unusual mechanism for the reduction of $\chi_{yyy}^{(2)}(0;0,0)$ from $t_{int}$=0 and $t_{int}$=$t_0$. **Fig.1b** and **Fig. S4c** [42] show the projections of $\chi_{yyy}^{(2)}(0;0,0)$ in the band structure and in the full BZ for the $t_{int}$=0 case, respectively. Interestingly, it is shown that the positive (negative) component of $\chi_{yyy}^{(2)}(0;0,0)$ is mostly contributed by the energy states around the *M* (*K*) point. A similar SHG component distribution is also found in the $t_{int}$=$t_0$ case, as shown in **Fig.1d** and **Fig. S4d** [42]. By calculating the SHG projection in the full BZ of $MoS_2/MoS_2$ under $t_{int}$=0 and $t_{int}$=$t_0$, we can obtain the difference between them. As shown in **Fig. 1e**, when the $t_{int}$ increases, the negative SHG component mostly increases around the $\Gamma$ point, consequently decreasing the $\chi_{yyy}^{(2)}(0;0,0)$ value. This finding can be further understood by the projection of $\chi_{yyy}^{(2)}(0;0,0)$ into each band at each *k* point in the band structure, as shown in **Figs. 1b** and **1d**. Interestingly, it is observed that this enhanced negative component of $\chi_{yyy}^{(2)}(0;0,0)$ from $t_{int}$=0 to $t_{int}$=$t_0$ is strongly associated with the increased band repulsion around $\Gamma$ point around the top valence band, which can effectively reduce the joint density of states (JDOS) at the band edge for optical transitions.

To further understand the role of $t_{int}$ on the $\chi_{yyy}^{(2)}(0;0,0)$, we have systemically calculated the $\chi_{yyy}^{(2)}(0;0,0)$ as a function of $t_{int}$ from 0 to 1.5$t_0$. As shown in **Fig. 1g**, it is seen that with the increase of $t_{int}$, the $\chi_{yyy}^{(2)}(0;0,0)$ will gradually decrease from $3.3\times10^4$ to $2.6\times10^4$ pm$^2$/V, a large reduction of ~21%. The three $d_{int}$ values corresponding to the different $t_{int}$ are also calculated and plotted in **Fig. 1g**. With the $t_{int}$ varying from 0.32$t_0$ to 1.0$t_0$ to 1.34$t_0$, the interlayer distance $d_{int}$ decreases from 4.26 Å to 3.0Å to 2.75 Å. Therefore, we can conclude that although the increase of $t_{int}$ can increase the TDMs, the increased band repulsion around $\Gamma$ point in the homo-bilayer MoS$_2$/MoS$_2$ can eventually decrease $\chi^{(2)}_{yyy}(0;0,0)$; the larger the $t_{int}$, the stronger band repulsion and consequently the smaller the $\chi^{(2)}_{yyy}(0;0,0)$ value.

### 3.2 $|\chi^{(2)}_{yyy}(-2\omega;\omega,\omega)|$ spectra of the homo-bilayer MoS$_2$/ MoS$_2$ as a function of $t_{int}$

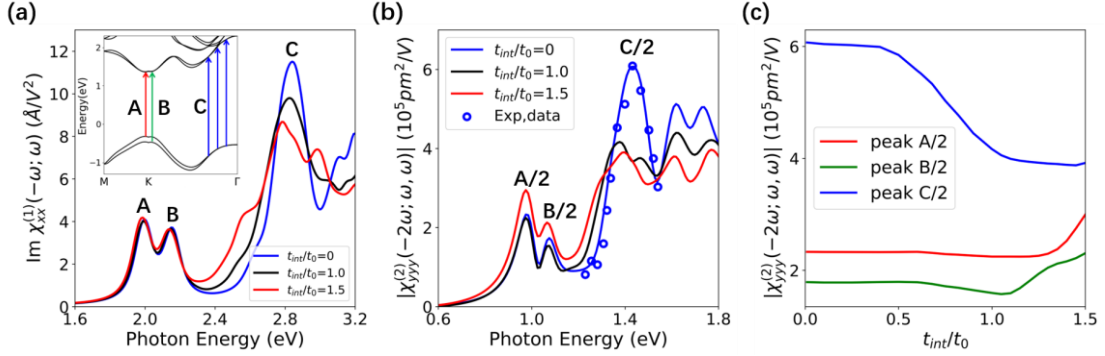

**Fig. 2.** (a) Linear $Im\chi^{(1)}(-\omega;\omega)$ (imaginary part of dielectric function) and (b) frequency-dependent SHG susceptibility $|\chi^{(2)}_{yyy}(-2\omega;\omega,\omega)|$ of the homo-bilayer MoS$_2$/ MoS$_2$ under different interlayer hopping strength $t_{int}$. Inset of (a): Arrows indicate the origin of optical transitions in the band structure contributed to A (A/2), B (B/2), and C (C/2) peaks in the $Im\chi^{(1)}(-\omega,\omega)$ ($|\chi^{(2)}_{yyy}(-2\omega;\omega,\omega)|$) spectra. In (b), blue dots indicate the experimentally measured values (enlarged by ×8) for the case of $t_{int}$=0, adopted from Ref.[12]. (c): Intensities of A/2, B/2, and C/2 peaks in the $|\chi^{(2)}_{yyy}(-2\omega;\omega,\omega)|$ spectra as a function of $t_{int}$.

Besides of the value of SHG susceptibility $\chi^{(2)}_{yyy}(0;0,0)$, it is interesting to further understand the $\omega$-dependent $\chi^{(2)}_{yyy}(-2\omega;\omega,\omega)$ as a function of $t_{int}$. Here we focus on the $|\chi^{(2)}_{yyy}(-2\omega;\omega,\omega)|$ spectrum, which is directly associated with the experimentally measured SHG intensities at different $\omega$. Generally, the linear and nonlinear optical spectra are strongly associated. To demonstrate the relationship between $Im\chi^{(1)}(-\omega;\omega)$ (imaginary part of dielectric function) and $|\chi^{(2)}_{yyy}(-2\omega;\omega,\omega)|$ under different $t_{int}$, both are calculated. **Fig. 2a** shows the calculated $Im\chi^{(1)}(-\omega;\omega)$ spectra for the MoS$_2$/MoS$_2$. There are three main peaks between 1.8 eV and 3.0 eV, which are marked as A, B, and C, respectively. As shown in the inset of **Fig. 2a**, the A and B peaks are induced by the optical transitions around the band edge at the *K* point, while the C peak originates from the strong band nesting effect along part of the *K* -*Γ* line [54]. Interestingly, when the $t_{int}$ increases, the intensities of A and B peaks in $Im\chi^{(1)}(-\omega;\omega)$ are almost unchanged but the intensity of C peak decreases significantly. This is mostly due to the change of $t_{int}$ mostly influences the band structures around the *Γ* point rather than the *K* point, as shown in **Figs. 1a** and **1c**. The larger the $t_{int}$, the larger the band dispersion around *Γ* point, consequently, the smaller the JDOS contributed to the optical transitions (**Fig. S5a** [42]). **Fig. 2b** shows the calculated $|\chi^{(2)}_{yyy}(-2\omega;\omega,\omega)|$ as a function of $t_{int}$. The $|\chi^{(2)}_{yyy}(-2\omega;\omega,\omega)|$ involves both single-photon ($\omega$) and double-photon ($2\omega$) resonances. Especially, the energy region of *E*<1.8 eV (below the bandgap of $MoS_2/MoS_2$) solely belongs to the double-photon resonance while the energy region of *E*>1.8 eV (not shown here) belongs to the mix of both single-photon and double-photon resonances. Here we focus on the low photon energy region of *E*<1.8 eV. Therefore, the A/2, B/2, and C/2 peaks shown in Fig. 2b all belongs to the pure $2\omega$ resonance. Since the energy positions of A/2, B/2 and C/2 peaks in the $|\chi^{(2)}_{yyy}(-2\omega;\omega,\omega)|$ spectra are exactly the half values of the corresponding A, B, and C peaks in the $Im\chi^{(1)}(-\omega;\omega)$ spectra, they should have the similar origin of optical transitions in the band structure (inset of **Fig. 2a**). As shown in **Fig. 2b**, the spectral position and shape of C/2 peak of bilayer $MoS_2/MoS_2$ ($t_{int}$=0) at ~ 1.45 eV fits well with the experimentally measured monolayer $MoS_2$ (marked in blue dots) [12]. We notice that our calculated C/2 peak intensity is roughly ×3.7 times larger than the experimentally measured one. This different, as also observed in other study [5,55], may be caused by either some uncertainties in the experiments (*e.g.*, the effect of substrate where substrate phonons or strain may be important) or the underestimation of the smearing factor δ in our calculations, which is related to the electronic relaxation time in the real samples. There have been experimentally measured SHG signals of A/2 and B/2 peaks for $MoS_2/ MoS_2$ bilayer [26]. Although due to different units, a direct comparison between our calculated values and the experimental data is unlikely, the calculated spectral positions and shapes of A/2 and B/2 peaks are similar to the experimentally measured ones for the decoupled bilayer $MoS_2$, as shown in **Fig. S6** [42]**.** As shown in **Fig. 2c**, the intensities of A/2 and B/2 peaks will only slightly change as a function of $t_{int}$. For example, the peak value of A/2 will keep almost unchanged in the region of 0<$t_{int}$<0.6$t_0$, and then slightly decreases from $2.3\times10^5$ to $2.2\times10^5$ pm$^2$/V in the region of $0.6t_0<t_{int}<1.2t_0$, and further increases from $2.2\times10^5$ to $3.0\times10^5$ pm$^2$/V in the region of $1.2t_0<t_{int}<1.5t_0$; however, the intensity of C/2 will largely decrease from $6.0\times10^5$ to $3.9\times10^5$ pm$^2$/V as the $t_{int}$ increase from 0 to $1.5t_0$, a strong reduction of ~35%. Our study can well explain the experimental observations that the SHG intensities of A/2 and B/2 peaks in the homo-bilayer $MoS_2/MoS_2$ can only slightly reduce from the decoupled case (inserted by $SiO_2$ thin-film, $t_{int}\sim0$) to the equilibrium case ($t_{int}=t_0$) [26].

We emphasize that, in practice, the exciton effects need to be considered using the accurate *GW+BSE* method, in order to capture the realistic optical spectra [15,55]. Fortunately, the underestimation of bandgap in the DFT-PBE calculations can accidentally obtain the similar peak positions as the experimental measurements [27]. In fact, there is one pioneering study [55] on discussing the influence of exciton effects on SHG response on monolayer $MoS_2$, using a tight-binding band structure and implementation of excitons in a Bethe-Salpeter framework. According to their findings: (1) the width of A/2 and B/2 peaks (**Fig. 2b**) in the SHG spectrum becomes narrower after the inclusion of exciton effects and the intensity of B/2 peak is slightly enhanced; (2) due to the cancellation of exciton effects and quasiparticle energy shifts, the A/2 and B/2 peak positions are indeed almost unchanged after the inclusion of exciton effects; (3) in the high photon energy region, e.g., larger than the C/2 peak position, excitonic effects can alter the resonance structure of the SHG spectrum.

### 3.3 $\chi^{(2)}_{yyy}(0;0,0)$ of the hetero-bilayer $MoS_2$/ $MoSe_2$ as a function of $t_{int}$

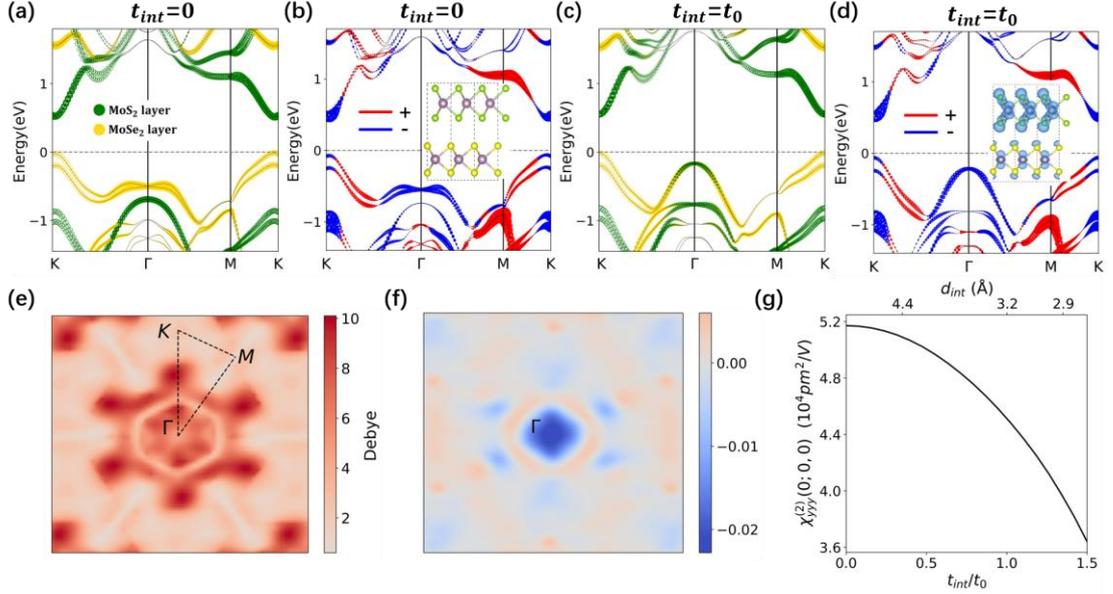

**Fig. 3.** (a) and (b) are orbital-resolved and SHG-weighted band structures of the hetero-bilayer MoS$_2$/MoSe$_2$ ($t_{int}$=0), respectively. Spin-orbital coupling (SOC) effects are considered in the band structure calculations. Fermi level is set to zero. Inset of (b): Side view of the atomic structure of MoS$_2$/MoSe$_2$. (c)-(d) are same as (a)-(b) but for the case of MoS$_2$/MoSe$_2$ ($t_{int}$=$t_0$). Inset of (d): SHG-weighted charge density of the MoS$_2$/MoSe$_2$ in the real space. See the enlarged versions of the insets in (b) and (d) in Fig. S2 [42]. (e): Difference of TDMs between the cases of $t_{int}$=0 and $t_{int}$=$t_0$ in the full BZ for the MoS$_2$/MoSe$_2$. Here, the plotted TDMs are the sum of all the four optical transition processes between the top two valence bands and the bottom two conduction bands. (f): Difference of SHG projection between the cases of $t_{int}$=0 and $t_{int}$=$t_0$ for the MoS$_2$/MoSe$_2$ in the full BZ. (g): Values of SHG susceptibility $\chi^{(2)}_{yyy}(0;0,0)$ for the MoS$_2$/MoSe$_2$ as a function of $t_{int}$.

The structure of hetero-bilayer MoS$_2$/MoSe$_2$ in the *3R* phase is shown as inset of **Fig 3b**. Due to the lattice mismatch between MoS$_2$ and MoSe$_2$, the lattice constant of MoS$_2$ (MoSe$_2$) is expanded (compressed) by ~2.8% (~2.1%) to build the small supercell for the convenience of our calculations. For the MoS$_2$/MoSe$_2$, due to the different energy levels of anion atomic orbitals, MoS$_2$ and MoSe$_2$ can form a type-II band alignment, giving rise to a largely reduced bandgap compared to its monolayer components. As shown in **Figs. 3a** and **3c**, the $E_g$ of MoS$_2$/MoSe$_2$ are 0.52 and 0.56 eV for the cases of $t_{int}$=0 and $t_{int}$=$t_0$, respectively. This indicates that the interlayer coupling has little effect on the bandgap values of MoS$_2$/MoSe$_2$, differing from the case of MoS$_2$/MoS$_2$ where a direct-to-indirect bandgap transition occurs. The band-edge states of MoS$_2$/MoSe$_2$ are mainly contributed by the Mo $d$ orbitals and Se $p$ orbitals. Similar to the case of bilayer MoS$_2$/MoS$_2$, we can see that the effect of $t_{int}$ on the band structure mainly reflects in the top valence band around the $\Gamma$ point, originated from the coupling between Mo $d_{z^2}$ and Se/S $p_z$ orbitals between the two neighboring layers. When the $t_{int}$ increases, as a type-II semiconductor, it is expected that the interlayer optical transitions in the MoS$_2$/MoSe$_2$ can be significantly increased, which is well confirmed by the calculated TDMs between the top two valence bands and the bottom two conduction bands in the full BZ from $t_{int}$=0 (**Fig. S7a** [42]) to $t_{int}$=$t_0$ (**Fig. S7b** [42]). As shown in **Fig. 3e**, we can observe that the optical transitions are largely increased around the $\Gamma$ point in the BZ.

Due to the C$_{3v}$ point group symmetry, besides of the dominated SHG susceptibility element $\chi^{(2)}_{yyy} = -\chi^{(2)}_{yxx} = -\chi^{(2)}_{xyx} = -\chi^{(2)}_{xxy}$, more nonzero SHG elements, i.e., $\chi^{(2)}_{zzz}$, $\chi^{(2)}_{zyy} = \chi^{(2)}_{zxx}$, $\chi^{(2)}_{xxz} = \chi^{(2)}_{yyz}$, appear in the MoS$_2$/MoSe$_2$. To simplify our discussion and also to compare with the case of MoS$_2$/MoS$_2$, we only present the $\chi^{(2)}_{yyy}$ in this study. It is found that, opposite to the increased TDMs, the calculated $\chi^{(2)}_{yyy}(0;0,0)$ of bilayer MoS$_2$/MoSe$_2$ decreases from 10.9×10$^4$ to 9.7×10$^4$ pm$^2$/V when $t_{int}$ increases from 0 to $t_0$. The SHG-weighted charge density (inset of **Fig. 3d**) shows that the $\chi^{(2)}_{yyy}(0;0,0)$ is mainly contributed by Mo $d_{x^2-y^2} + d_{xy}$, Mo $d_{z^2}$, and S $p_z$ orbitals, regardless of the specific value of $t_{int}$.

Similar to the bilayer MoS$_2$/MoS$_2$, as shown in **Figs. 3b** and **3d** (also see **Figs. S7c** and **S7d** for the SHG projection in the full BZ [42]), the positive and negative components of $\chi^{(2)}_{yyy}(0;0,0)$ in the MoS$_2$/MoSe$_2$ are mainly contributed by the electronic states around $M$ and $K$ points, respectively. As shown **Fig. 3f**, when the $t_{int}$ increases, the interlayer orbital coupling can significantly increase the negative component of $\chi^{(2)}_{yyy}(0;0,0)$, reducing the $\chi^{(2)}_{yyy}(0;0,0)$ value. As shown in **Fig 3g**, the calculated $\chi^{(2)}_{yyy}(0;0,0)$ value gradually decreases from 5.2×10$^4$ pm$^2$/V ($t_{int}$=0) to 3.6×10$^4$ pm$^2$/V ($t_{int}$=1.5$t_0$), a large reduction of ~31%. Three $d_{int}$ values corresponding to the different $t_{int}$ are also calculated and marked in **Fig. 3g**. With the $t_{int}$ increases from $0.35t_0$ to $1.0t_0$ to $1.35t_0$, the $d_{int}$ decreases from 4.4 Å to 3.2 Å to 2.9 Å. Therefore, similar to the homo-bilayer $MoS_2/MoS_2$, we can conclude that although the increase of $t_{int}$ can increase the TDMs, the increased band repulsion around the $\Gamma$ point in the hetero-bilayer $MoS_2/MoSe_2$ can still decrease $\chi^{(2)}_{yyy}(0;0,0)$; the larger the $t_{int}$, the stronger band repulsion around the $\Gamma$ point and consequently the smaller the $\chi^{(2)}_{yyy}(0;0,0)$.

### 3.4 $|\chi^{(2)}_{yyy}(-2\omega;\omega,\omega)|$ spectra of the hetero-bilayer $MoS_2/MoSe_2$ as a function of $t_{int}$

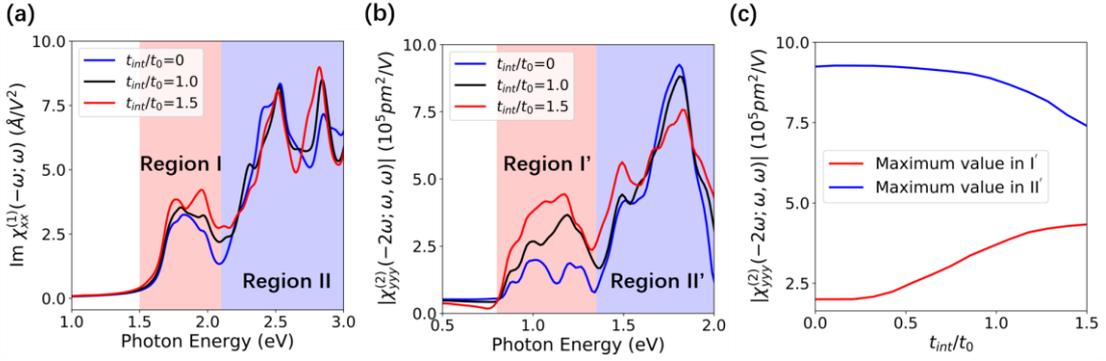

**Fig. 4.** (a) Linear $Im\chi^{(1)}(-\omega;\omega)$ (imaginary part of dielectric function) and (b) frequency-dependent SHG susceptibility $|\chi^{(2)}_{yyy}(-2\omega;\omega,\omega)|$ of the hetero-bilayer $MoS_2/MoSe_2$ with different interlayer hopping strength $t_{int}$. (c): Maximum values of the $|\chi^{(2)}_{yyy}(-2\omega;\omega,\omega)|$ spectra in the region I' and region II' (marked in (b)) as a function of $t_{int}$.

**Fig. 4a** shows the calculated linear $Im\chi^{(1)}(-\omega;\omega)$ spectra for the $MoS_2/MoSe_2$. Generally, there are multiple peaks in the low-photon-energy region I (1.5~2.1 eV) and the high-photon-energy region II (2.1~3.0 eV). When $t_{int}$=0, without the interlayer interactions, the peaks in the region I are mostly contributed by the band-edge optical transitions around the $K$ point in each individual intralayer, while the interlayer optical transitions are completely forbidden (see **Fig. S8a** [42]). When $t_{int}$≠0, the interlayer optical transitions around the band edge appear and increase as the $t_{int}$ increases (see **Fig. S8b** [42]), giving rise to the enhanced optical transitions in the region I of $Im\chi^{(1)}(-\omega;\omega)$. On the other hand, when $t_{int}$=0, the multiple peaks with strong intensities in the region II are mostly contributed by the optical transitions around the $\Gamma$ point and the band nesting effect along part of the $\Gamma$-$K$ line in each individual intralayer. When $t_{int}$≠0, the interlayer optical transitions appear in the region II and increase as the $t_{int}$ increases;

meanwhile, the band nesting effect is reduced due to the increased interlayer orbital coupling. Therefore, the coexistence of these two completing effects give rise to a complicated evolution of these peaks in region II as a function of $t_{int}$.

We have further calculated the $|\chi^{(2)}_{yyy}(-2\omega;\omega,\omega)|$ spectra for three different $t_{int}$ values, as shown in **Fig. 4b**. Differing from the case of homo-bilayer $MoS_2/MoS_2$ (**Fig. 2b**), it is difficult to distinguish the contributions from single-photon and double-photon resonances in these spectra. Interestingly, the multiple peaks in $|\chi^{(2)}_{yyy}(-2\omega;\omega,\omega)|$ spectra can also be divided into the two regions, *i.e.*, low-phonon-energy region I' (0.8~1.3 eV) and high-phonon-energy region II' (1.3~2.0 eV). When $t_{int}$ increases, the interlayer optical transitions can significantly increase the intensities of peaks in region I'. As shown in **Fig. 4c**, the maximum $|\chi^{(2)}_{yyy}(-2\omega;\omega,\omega)|$ in region I' gradually increases from $2.0\times10^5$ to $4.2\times10^5$ pm$^2$/V, a huge increase of ~110 %. Differing from the region I', the changes of $|\chi^{(2)}_{yyy}(-2\omega;\omega,\omega)|$ in the high-photon-energy region II' could be mainly contributed by the intralayer optical transitions. Especially, the SHG peak at ~1.8 eV is caused by the band nesting effect along part of the *K-Γ* line (**Fig. S5b** [42]). The larger the $t_{int}$, the weaker the band nesting effect, and consequently the weaker the $|\chi^{(2)}_{yyy}(-2\omega;\omega,\omega)|$ value at ~1.8 eV. As shown in **Fig. 4c**, in contrast to the region I, the maximum $|\chi^{(2)}_{yyy}(-2\omega;\omega,\omega)|$ value in region II' (i.e., peak value at ~1.8 eV) will gradually decrease from $9.2\times10^5$ to $7.4\times10^5$, a small reduction of ~20%. Interestingly, we emphasize that since the band nesting effect in the hetero-bilayer $MoS_2/MoSe_2$ (**Fig. S5b** [42]) is weaker than that in the homo-bilayer $MoS_2/MoS_2$ (**Fig. S5a** [42]), the change of SHG peak intensity is also less noticeable in $MoS_2/MoSe_2$ (~20%) than in $MoS_2/MoS_2$ (~35%).

Recently, there are experimental studies on the effects of interlayer coupling on the SHG intensity in 3R-type hetero-bilayer $MoS_2/MoSe_{2(1-x)}Se_{2x}$ samples [26]. Interestingly, it is found that there is a strong enhancement of SHG intensity at the low photon energy range (~1.0 eV) from the decoupled sample ($t_{int}$~0) to the coupled one ($t_{int}=t_0$), which is very different from the observations in the homo-bilayer $MoS_2/MoS_2$ samples. The physical origin behind these puzzling experimental observations in the different TMDs under variable $t_{int}$ can be well explained by our calculations.

## IV. CONCLUSION

In summary, taking homo-bilayer MoS$_2$/MoS$_2$ and hetero-bilayer MoS$_2$/MoSe$_2$ as typical examples, we have systemically investigated their SHG susceptibilities $\chi^{(2)}$ as a function of $t_{int}$ using first-principles Wannier function calculations. For the $\chi^{(2)}_{yyy}(0;0,0)$ of MoS$_2$/MoS$_2$ and MoS$_2$/MoSe$_2$, it is found that the larger the $t_{int}$, the smaller the $\chi^{(2)}_{yyy}(0;0,0)$. For $|\chi^{(2)}_{yyy}(-2\omega;\omega,\omega)|$ of MoS$_2$/MoSe$_2$ in the low photon-energy region, differing from the MoS$_2$/MoS$_2$, their peak values are very sensitive to the $t_{int}$; the larger the $t_{int}$, the larger the $|\chi^{(2)}_{yyy}(-2\omega;\omega,\omega)|$. For dynamical $|\chi^{(2)}_{yyy}(-2\omega;\omega,\omega)|$ of MoS$_2$/MoS$_2$ in the high photon-energy region, comparing to the MoS$_2$/MoSe$_2$, their peak values will gradually decrease in a much more noticeable way as the $t_{int}$ increases. Our study not only can successfully explain the recent experimental observations on the dramatically different SHG responses in different bilayer TMDs as a function of interlayer coupling strengths, but also may provide a general approach to tune the SHG effects in the vdW bilayer systems.

**Acknowledgements:** The authors thank J. F. Wang, Y. Li, and B. Shao for helpful discussions. This work is supported by the NSFC (Nos. 12088101, 11634003) and NSAF U1930402. Part of calculations are done in Tianhe-JK cluster at CSRC.